\documentclass{aa}

\usepackage{graphicx, epstopdf, multicol, blindtext}
\usepackage{txfonts, color, caption, subcaption, paralist}
\usepackage[normalem]{ulem}
\usepackage{natbib}
\bibpunct{(}{)}{;}{a}{}{,}
\begin{document}

   \title{Models for the evolution of close binaries with He-Star and WD components towards Type Ia supernova explosions}

   \author{P. Neunteufel
          \inst{1}
          \and
          S.-C. Yoon
          \inst{2}
          \and
          N. Langer
          \inst{1}
          }

   \institute{\inst{1}Argelander Institut f\"ur Astronomy (AIfA), University of Bonn,
              Auf dem H\"ugel 71, D-53121 Bonn\\
              \inst{2}Department of Physics and Astronomy, Seoul National University,
              599 Gwanak-ro, Gwanak-gu, Seoul, 151-742, Korea\\
              \email{neunteufel@astro.uni-bonn.de}}

   \date{Received (month) (day), (year); accepted (month) (day), (year)}
\abstract
   {Type Ia supernovae (SNe Ia) have been an important tool for astronomy for quite some time; however, the nature of their progenitors remains somewhat mysterious. Recent theoretical studies indicated the possibility of producing thermonuclear detonations of carbon-oxygen white dwarfs (CO WDs) at masses less than the Chandrasekhar mass through accretion of helium-rich matter, which would, depending on mass accretion rate, mass, and initial temperature of the WD, spectrally resemble either a normal SN Ia or a peculiar one.}
   {This study aims to further resolve the state of binary systems comprised of a sub-Chandrasekhar-mass CO WD and a helium star at the point where an accretion-induced detonation occurs and constrains the part of the initial parameter space where this kind of phenomenon is possible.}
   {Preexisting data obtained through simulations of single, constantly accreting CO WDs is used as an indicator for the behavior of new binary models in which the WD is treated as a point mass and which include the non-degenerate partner as a fully resolved stellar model. We parameterize the ignition of the accumulated helium layer, changes in the WD temperature, and changes in the CO core mass depending on the mass transfer rate.}
   {The initial conditions allowing for detonation do not form a single contiguous area in the parameter space, whose shape is profoundly influenced by the behavior of the donor star. Mass loss due to Nova outbursts acts in favor of detonation. According to our criteria, about 10\% of the detonations in this study can be expected to show spectra consistent with ordinary SNe Ia; the rest exhibit peculiar features.}
   {}

   
   \titlerunning{Evolution of Close Binaries with He-Star and WD Components}
   \maketitle
\section{Introduction}
 Several decades ago, sub-Chandrasekhar mass thermonuclear detonation of carbon-oxygen white dwarfs (CO WD) initiated by an explosive ignition in an accreted layer of helium were  deemed a promising model for a number of observed exploding stellar transients. These transients include Type Ia supernovae (SNe Ia) and other classes of transients resembling SNe Ia, like the spectrally peculiar ``Iax'' \citep[see][]{FCC2013} and, probably, subluminous ``point Ia'' (.Ia) Supernovae \citep{BSWN2007}. \cite{GHP2014} discuss the observational properties of peculiar SNe Ia, while observational evidence of the nature of the progenitors of SNe .Ia was presented by \cite{KHG2014}.

The possibility of a detonation in an accumulated helium envelope of a sub-Chandrasekhar mass CO WD was first investigated in the early 1980s \citep{T1980,T1980b,N1980P,N1982b,N1982a}. It was soon realized that, depending on a number of factors, ignition in the accumulated He-shell either led to the ignition and subsequent ejection of the He-shell only, leaving an intact CO core, or led to a secondary detonation of the CO core and thus the complete disruption of the WD. The mechanism leading to the complete destruction of the WD has become known as the ``double detonation'' scenario and has been investigated in a number of one- and multidimensional studies \citep{L1990, LG1990, LG1991, LA1995, B1997P, L1997P, FHR2007, FRH2010, SRH2010, KSF2010, WK2011}. This scenario is a subset of what has become known as ``WD+He star channel''.

This channel forms part of a picture of He-accreting WDs, introduced as candidates for SN Ia or SN .Ia progenitors, which also includes double degenerate systems featuring a He WD instead of a He star. Arguments in favor of the WD+He WD channel have been investigated by \cite{SB2009}.

During the course of the study of helium accreting WDs \citep{BSWN2007, FHR2007}, it was realized that helium detonations were possible even in comparatively low mass ($< 0.0035\,\text{M}_\odot$) envelopes of accumulated material and that a detonation in the helium envelope of the WD (characterized by a supersonic shock in the medium) will robustly detonate the CO core \citep{FRH2010, WK2011}.
It was also realized that synthetic spectra produced for some of the models investigated in the studies mentioned above resembled those of observed ``normal'' SNe Ia \citep{SRH2010, KSF2010, WK2011}, while others more closely resembled those of SNe Iax.

These findings stimulated interest in whether double detonations could make a significant contribution to the rate of observed SNe Ia and that their subluminous and spectrally peculiar kin. More recently, a study by \citet{WJH2013} used detailed binary evolution models and subsequent binary population synthesis calculations to investigate the rate of SNe Iax that might result from double detonations. In that study, under the assumption that double detonation would always lead to a SN Iax or, at the very  least, a spectrally peculiar SN, it was found that the rate of these phenomena can be adequately explained using the helium accretion double detonation model, assuming detonation at the point where a fixed amount of helium ($0.1\,\text{M}_\odot$) has  accumulated on the WD \citep[cited sources:][]{IT2004,RBS2011}. 
%Binary evolution calculations with variable ignition mass have been performed by \cite{ZWZ2014}.

Very recently, numerical studies using population synthesis simulations and the Cambridge stellar evolution code \citep{LMSW2015,LSAW2015} have suggested the WD+He star channel as a promising candidate mechanism for the creation of SNe Iax, arguing, however, that a small fraction of SN Iax may be the result of helium deflagrations (see Sec.~\ref{sec:methods}) on the WD component rather than detonation.

In reality, the amount of accumulated helium in a CO WD at detonation is expected to be a function of the initial mass and temperature of the CO WD and of the mass accretion rate of helium-rich matter. This has  been investigated by \cite{RBS2014}, who compared occurrence rates of helium accretion-induced detonations of white dwarfs using a variation of assumptions regarding input physics, including a prescription of variable shell ignition masses adapted from \cite{IT1989}.
Whether the helium detonation leads to a peculiar SN event or to an ordinary-looking SN Ia would also depend on this. In our study we combine some of the data provided by preexisting research \citep[mainly from][]{WK2011} with new binary evolution calculations to investigate the impact of allowing for the mass of the helium accumulated on the WD to vary with parameters such as the system's mass transfer rate and the mass of the accreting WD on the distribution of detonating systems in the parameter space. From this data, we generate a distribution of the expected helium shell masses, allowing for an estimate of the ratio of high to low final helium shell masses, which have an impact on the spectra produced by double detonation~\citep{SRH2010, WK2011}.  We are also able to make a statement about the post-SN state of the donor star, which might be of use in the explanation of single high velocity stars \citep[see][]{GFZ2015}.

This article is organized as follows. In Sec.~\ref{sec:methods} we describe the input physics of our models and the  computational framework. In Sec.~\ref{sec:results},  we first discuss the behavior of helium donor stars using single helium star models (Sec.~\ref{sec:donor}), and describe physical and numerical restrictions and their impact on our choice of the considered parameter space.  Then (Sec.~\ref{sec:binaryresults}) we present binary evolution models for which mass transfer from the helium donor star is computed self-consistently with the orbital evolution of the binary system, while the CO WD is approximated as a point mass, and discuss the observational implications. We conclude our study in Sec.~\ref{sec:conclusions}. 

\section{Numerical methods and physical assumptions} \label{sec:methods}

\subsection{Detonation conditions and mass transfer efficiency}

\begin{figure}
   \centering
   \small
   \input{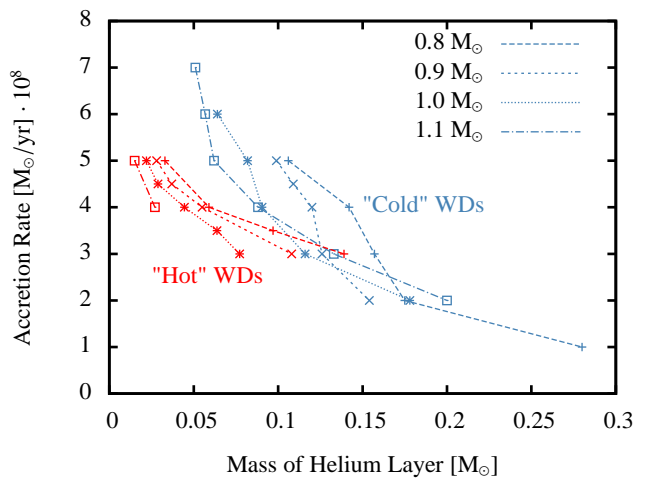}
   \normalsize
      \caption{Points of helium detonation as predicted by \cite{WK2011} in the relevant mass range. Points and lines plotted in blue are valid for cold dwarfs, red points and lines for hot dwarfs. Masses are given in units of $\text{M}_{\odot}$. We assume that a detonation on the white dwarf component  occurs when a system crosses the interpolated lines.}
         \label{fig:detlines}
\end{figure}

The ability of single white dwarfs undergoing accretion of He-rich matter at a constant rate to produce a detonation in the accumulated He-shell has been studied in some detail by \cite{WK2011}. A number of cases characterized by the WD mass and the mass accretion rate were simulated and many of them produced helium detonations powerful enough to detonate the CO core either by way of an inward shock, if the detonation occurred above the surface of the CO core, or by simple compression, if the detonation occurred at the surface of the CO core.
For CO white dwarfs between $0.7\,\text{M}_{\odot}$ and $1.1\,\text{M}_{\odot}$, a helium ignition violent enough to lead to CO core detonation may occur at constant accretion rates $\dot{M}$ between $1\cdot 10^{-8}\,\text{M}_{\odot}/\text{yr}$ and $5\cdot 10^{-8}\,\text{M}_{\odot}/\text{yr}$. The critical mass of the helium shell depends on the initial temperature and mass of the white dwarf for a given mass accretion rate. The initial temperatures of their models are defined according to the initial luminosity: hot with $L = 1.0\,\text{L}_{\odot}$ and cold with $L = 0.01\,\text{L}_{\odot}$. Accretion rates and final helium shell masses for detonating models as found by \cite{WK2011} are shown in Fig.~\ref{fig:detlines}.
It has also been shown that helium ignition, if occurring at accretion rates higher than those indicated in Fig.~\ref{fig:detlines}, would not lead to a subsequent detonation of the CO core. These systems instead underwent a ``deflagration'' event characterized by a subsonically propagating burning front, which led to the ejection of the shell in a nova-like process, leaving the CO core intact. Detonations, however, always led to a subsequent detonation of the CO core, provided the models' zoning was sufficiently fine.

\begin{figure}
        \centering
        \small
       \input{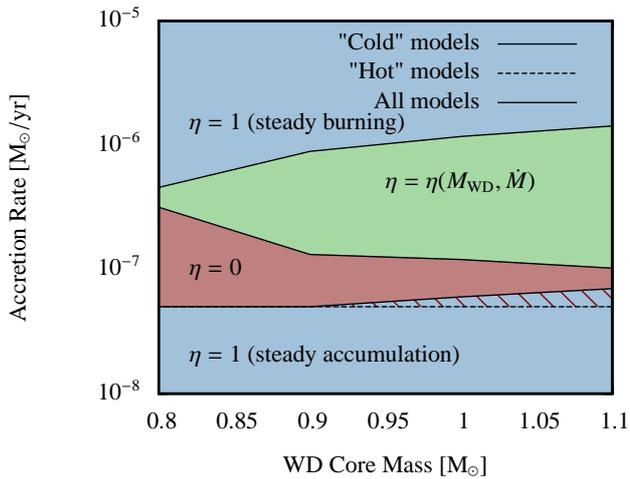}
        \normalsize
                \caption{Illustration of the parameters of the prescription used for the mass accumulation efficiency (see Eq.~\ref{eq:acceff_prescription}) at different mass accretion rates. In the area denoted by red-blue stripes, the upper limit for $\eta = 0$ differs for hot and cold models. The dash-dotted line applies to cold models and the dashed line to hot models. $\eta$ is changed from $1$ to $0$ when the time-averaged accretion rate crosses the dash-dotted or dashed lines, while for all other lines the actual accretion rate is the deciding parameter.
Note that the upper $\eta=1$-regime denotes steady helium burning while the lower one denotes steady accumulation of helium on the CO core without nuclear burning.}
                \label{fig:acceff_prescription}
\end{figure}

The response of the WD to accretion is described in terms of the the amount of mass retained on the WD versus the amount of mass accreted onto the white dwarf (with the remaining mass being ejected from the system). This relation is called ``mass acculumation efficiency'', which can be expressed as
\begin{equation}
\eta = \dot{M}_\text{acc} / \dot{M}~,
\end{equation}
where $\dot{M}_\text{acc}$ is the rate of material accumulated on the accretor and $\dot{M}$ is the rate of material being transferred from the donor.
Depending on the mass transfer rate, in the case of $\eta>0$ the retained material either simply accumulates in the form of a helium layer of increasing thickness (low mass transfer rates), hereafter referred to as ``steady accumulation'' or is partially or fully processed into carbon or oxygen (higher mass transfer rates).
At accretion rates higher than those allowing for steady accumulation, consecutive nova outbursts preclude the buildup of a helium layer of sufficient mass to trigger a CO core detonation. At even higher accretion rates, a stable nuclear burning shell may develop on the WD, leading to steady growth of the CO core.
The value of $\eta$ depends on the the mass of the white dwarf and the helium accretion rate.\\
The  evolution of $\eta$ for different white dwarf masses and accretion rates has been studied by \cite{HK2004}, whose estimates we use whenever our systems show accretion rates high enough to make them a viable option. Numerically, the resulting values for $\eta$ fall between $0.3$ and $1$ increasing with the mass transfer rate. We use the following prescription for $\eta$ over the whole range of occurring mass transfer rates (see also Fig.~\ref{fig:acceff_prescription}),
\begin{equation} \label{eq:acceff_prescription}
  \eta = \begin{cases}
    1, & \text{if $0 < [\dot{M}] < \dot{M}_{\text{WK,max}}$}~,\\
    0, & \text{if $\dot{M}_{\text{WK,max}} < \dot{M} < \dot{M}_{\text{KH,min}}$}~,\\
    \eta_{KH}(M_{\text{WD}},\dot{M}), & \text{if $\dot{M}_{\text{KH,min}} < \dot{M}$}~,\\
    1, & \text{if $\dot{M}_{\text{KH,max}} < \dot{M}$,}
  \end{cases}
\end{equation}
where $\dot{M}_{\text{WK,max}}$ is the upper limit of the mass transfer rates studied by \cite{WK2011} corresponding to the maximum sustainable mass transfer rate for steady accumulation, $\dot{M}_{\text{KH,min}}$ the lower limit of the range studied by \cite{HK2004} corresponding to the minimum sustainable rate for $\eta \geq 0$, and $[\dot{M}]$ is the time averaged mass transfer rate.

\subsection{Computational framework}

The time-dependent evolution of the donor star properties, mass transfer rate, and binary orbital separation are calculated using the binary evolution code (BEC), a well-established simulation framework capable of performing detailed one-dimensional experiments of single or binary systems \citep{LDWH2000, YL2004}. This framework resulted from a development by \cite{B1998}  of a preexisting implicit hydrodynamic stellar evolution code \citep{L1991, L1998}. Here we study short-period binary systems with initial orbital periods ranging from $0.02$ to $0.12$ days, WD masses between $0.8$ and $1.0\,\text{M}_{\odot}$, and helium star donor masses between $0.5\,\text{M}_{\odot}$ and $1.0\,\text{M}_{\odot}$. The choice of the boundaries to the initial parameter space is explained in greater detail in Sec.~\ref{sec:parameterspace}. The WD component is approximated as a point-mass, while the He star component is a fully resolved, non-rotating, non-magnetic stellar model at solar metallicity. \\
The BEC code incorporates angular momentum loss due to gravitational wave radiation (GWR; see \citealt{LL1975} for a mathematical description).
Mass transfer is calculated by solving implicitly the equation
\begin{equation}
R_L - R + H_P \ln \left( \frac{\dot{M}}{\dot{M}_0} \right) = 0~
\end{equation} 
\citep[see][]{R1988, RK1990}. 
Here, $H_P$ is the photospheric pressure scale height, $R$ the stellar radius as defined by the lower edge of the photosphere, $R_L$ the Roche lobe radius for which we use the approximation by \cite{E1983}, and 
\begin{equation}
\dot{M}_0 = \frac{1}{\sqrt{e}} \rho v_S Q~,
\end{equation}
where $v_S$ is the speed of sound in a plasma as defined by $v_S^2 = \Re T/\mu$ with $\Re$ being the ideal gas constant, $T$ the plasma temperature,  $\mu$ the mean molecular weight, and $Q$ is the effective stream cross section calculated as prescribed by \cite{MH1983}.

At this point it should be noted that WDs in the mass range of $M_\text{WD} = 0.8\,-\,1.0 \text{M}_\odot$ take about 0.2 Gyr to cool from $L = 1.0\,\text{L}_{\odot}$ to $L = 0.01\,\text{L}_{\odot}$ \citep{RAM2010}. This means that the shortest period systems would not have sufficient time to cool to $L = 0.01\,\text{L}_{\odot}$. Therefore, the temperature of the white dwarf at the end of the common envelope phase necessary to produce this kind of system will have a significant impact on the subsequent binary interaction. Detailed investigations of the thermal evolution of a white dwarf during a common envelope phase would be required to allow more precise predictions. However, such studies have not found their way into the literature as of this time.

Systems in which $\dot{M}$ exceeds $\dot{M}_{\text{WK,max}}$ but remains in the area of the parameter space where $\eta<1$ are in a phase of consecutive nova outbursts. 
If the mass of the CO core increases with stable helium burning, the corresponding detonation line is obtained by interpolating the existing detonation lines (Fig.~\ref{fig:detlines}). In the case of the WD mass exceeding $1.1\,\text{M}_{\odot}$, which is the largest considered by \cite{WK2011}, our simulation is aborted since the non-monotonic  nature of the detonation points with respect to the CO core mass makes simple extrapolation towards higher masses a risky option. The great majority of systems containing donors more massive than $1.0\,\text{M}_{\odot}$ lead to CO cores heavier than $1.1\,\text{M}_{\odot}$ because the nuclear timescales of the heavy donors lead to high enough mass transfer rates to induce stable burning of the accreted material. This increases the CO core mass considerably and precludes detonation via the double detonation mechanism. This is the reason why we limit the parameter space considered in this study to donor and WD masses smaller than $1.0\,\text{M}_{\odot}$.\\
During the research described in this work, estimates of a wider range of parameters of the  reactions to and outcomes of mass accretion onto WDs has become available, and any future study of this kind would be well advised to utilize them \citep{PTY2014}. We would also like to emphasize that the accretion efficiencies given by \cite{HK2004} are somewhat high and that the discrepancy with \cite{PTY2014} has been shown to have a non-negligible influence on the expected event rates and on the  resulting final He-shell masses \citep{RBS2014}. Crossover from the lower $\eta=1$-regime into the $\eta=0$-regime is treated as an indicator that the system undergoes a deflagration (i.e., a subsonic shell explosion), during which all of the accumulated helium is ejected, with the CO core left intact. If the WD model is in a ``cold'' state when this crossover occurs, it is switched to a ``hot'' state for the remainder of the simulation.

The resulting sudden loss of mass in a Nova event has a significant impact on the Roche lobe radii, which heavily influences the system's mass transfer rate, and therefore has to be taken into account. This is accomplished by subtracting the mass of the helium shell from the mass of the white dwarf whenever the time averaged mass transfer rate exceeds the range given in Fig.~\ref{fig:detlines}. Since the binary evolution code assumes circular orbits for binary systems, the angular momentum loss associated with the loss of mass from the accretor is dealt with by changing the orbital separation to that of a circular orbit of the same angular momentum as the realistic, elliptical orbit after mass loss.

\section{Simulation results} \label{sec:results}
\subsection{Evolution of the donor star} \label{sec:donor}

\begin{figure*}
   \centering
   \small
   \input{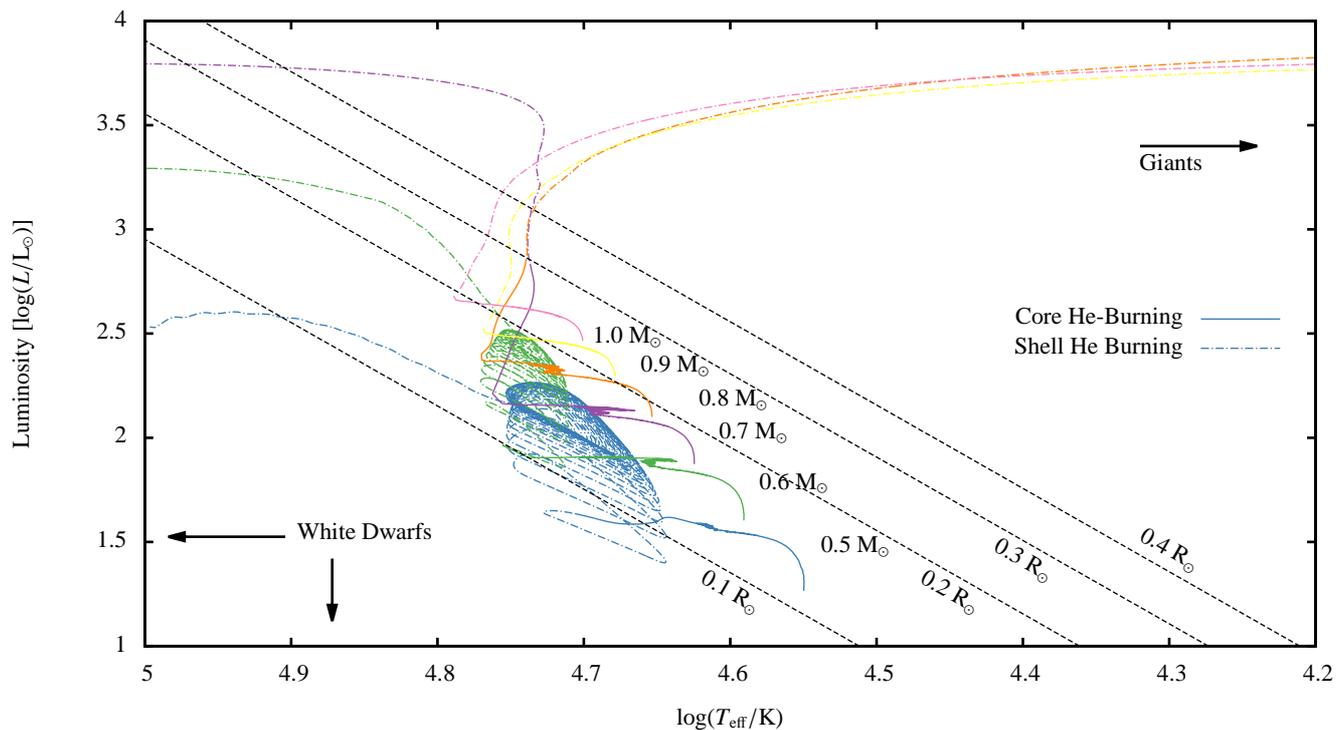}
   \normalsize
      \caption{Evolutionary tracks of single He stars with masses as used in this study in the HR-diagram. A large number of thermal pulses occur in the three lowest mass models after the end of core helium burning.}
         \label{fig:singlehd}
\end{figure*}

As described in Sec.~\ref{sec:methods}, in this study we treat the white dwarf component of the system as a point mass. Its state is entirely defined by the mass of the white dwarf and the mass of the helium layer that has been deposited on it, and by the system's current mass transfer rate. This means that the evolution of the system is dominated by the evolution of the orbital separation, which is a product of the history of the binary interaction and  of angular momentum loss due to GWR, and the evolution of the He-star. The evolution of the donor star is influenced by mass transfer. In this section, we try to disentangle those behaviors that would be exhibited by the donor star in isolation from those that result from binary interaction.

The evolution of single helium stars has been studied extensively in the past \citep[see, e.g.,][]{E2006}, but in the interest of consistency, we discuss below the evolution of the particular models used in this study. 
These models  were created by following the evolution of a hydrogen main sequence star up to the onset of helium burning and then removing the remaining hydrogen envelope. The mass of the resulting model was then self-consistently changed to the desired value and, with nuclear energy generation switched off, computed until thermal equilibrium was achieved.
Fig.~\ref{fig:singlehd} shows the theoretical paths through the Hertzsprung-Russel diagram (HR-diagram) taken by He stars in isolation. All of these stars possess convective cores having $M_\text{core}/M_\text{star} = 0.2$ in our $0.5\,\text{M}_\odot$ models and $M_\text{core}/M_\text{star} = 0.25$ in the $1.0\,\text{M}_\odot$ models at the helium zero age main sequence, which grow to between $M_\text{core}/M_\text{star} = 0.3$ and $M_\text{core}/M_\text{star} = 0.55$ at the end of core helium burning, respectively. Thermal pulses develop for $0.5\,\text{M}_\odot$ -- $0.7\,\text{M}_\odot$ after the core helium burning phase with radius changes of a factor of 2 to 3. Any mass transfer occurring during these pulses will exhibit mass transfer rates according to the thermal timescale ($\dot{M}\sim 1\cdot10^{-6} \text{M}_\odot/\text{yr}$), which is above the regime required for detonation (see Fig.~\ref{fig:detlines}). After the end of the pulse phase, all of the low mass stars will become CO WDs. 

The more massive He stars ($>0.7\,\text{M}_\odot$) exhibit no significant pulses, but after the end of core helium burning they  expand to become giants. At this point the simulation is stopped with the understanding that mass transfer rates during the giant phase of a donor would be far too high for double detonation to be possible. These systems would most likely enter a common envelope phase. In any case, these systems lie outside the parameter space accessible to this study (see Sec.~\ref{sec:parameterspace}).

\subsection{Initial parameters of the investigated binary systems} \label{sec:parameterspace}

The parameter space under scrutiny in this study covers initial white dwarf masses $M_\text{WD} = 0.8\,\text{M}_\odot ~ ... ~ 1.0_\odot$, $M_\text{He} = 0.5\,\text{M}_\odot ~ ... ~ 1.0\,\text{M}_\odot$, and initial orbital periods $P_\text{init} = 0.03 ~ ... ~ 0.11 ~ \text{days}$.
Since we use the results of \cite{WK2011} for this study, we are generally restricted to the parameter space of white dwarf masses investigated by them ($0.7\,\text{M}_\odot \leq M_\text{WD} \leq 1.1\,\text{M}_\odot$). Their data for $M_\text{WD}=0.7\,\text{M}_\odot$ is too sparse for us to use. Their data for $M_\text{WD}=1.1\,\text{M}_\odot$ is less sparse, but since the CO core masses of our white dwarfs may grow over the course of our calculations, we would be forced to extrapolate the existing data to CO core masses beyond $1.1\,\text{M}_\odot$. 
This also limits the range of donor masses accessible to us. Since the nuclear timescale of helium stars of masses $M_{\text{Donor}}>1.0~\text{M}_\odot$ approaches $10^6~\text{yr}$, mass transfer will be in the steady burning regime ($\eta=1$) already on the helium main sequence \citep[see also][]{YL2003}. Calculations performed during the course of this study have shown that in these cases the CO core mass of the accretor invariably exceeds $1.1~\text{M}_\odot$ at some point, regardless of the chosen initial WD mass or initial orbital period.
We thus set the upper limit for the range of donor star masses at the cited value of $M_{\text{Donor}}=1.0~\text{M}_\odot$. \\
The lower boundary of our donor star mass range is provided by the inherent behavior of low mass helium stars. Helium stars with masses below $0.3\,\text{M}_\odot$ fail to ignite altogether, and contract to become He white dwarfs. Helium stars with masses of $0.4\,\text{M}_\odot$  experience nuclear burning in their core and expand, resulting in Roche-lobe overflow (RLOF) in binary systems, but owing to their long nuclear timescale ($\sim 10^8 \text{yr}$), a rather large proportion (more than 50~\% in all cases) of their initial mass needs to be lost to the companion in order to initiate a detonation. These models usually fail to converge after a significant amount of mass has been lost to the accretor. The donor, however, is expected to eventually become a CO WD. Readers are referred to \cite{PYT2015} and \cite{BBM2015} for detailed evolutionary models of these systems. \\
The lower end of the initial period distribution is provided by the point for which the radius of a donor star at helium zero age main sequence equals that of its Roche-lobe. The upper end is the point where the lowest mass donor stars do not experience RLOF before becoming electron degenerate. Since higher mass donors become giants for periods larger than $0.12~\text{d}$, low mass transfer rates that lead to He detonation are not expected. This leads us to the conclusion that periods longer than $0.12~\text{d}$ are not relevant to the present study. \\
We expect our calculations to retain some predictive power when moving to smaller WD masses, provided their composition is similar. However, \cite{IT1985} showed that some possible outcomes of binary main sequence evolution contain hybrid HeCO white dwarfs as an intermediate between the mass ranges for archetypical He WDs and CO WDs. Whether the results in these cases are comparable would have to be verified independently.

\subsection{Discussion of binary results} \label{sec:binaryresults}

\subsubsection{Mass accretion modes}

The most obvious information to be gained from this study is which systems in the initial parameter space might result in a detonation. For the discussions that follow, the initial parameter space where detonation occurs is called the detonation zone. The graphs depicted in Fig.~\ref{fig:grids} indicate the positions of all systems in this study according to the initial mass of the WD and their positions in the initial parameter space. Furthermore, the color shading indicates the amount of helium accumulated until detonation.
Our calculations have allow us to distinguish four different kinds of systems:

\begin{figure*}
   \centering
   \small
   \begin{subfigure}[t]{1.0\textwidth}
   \input{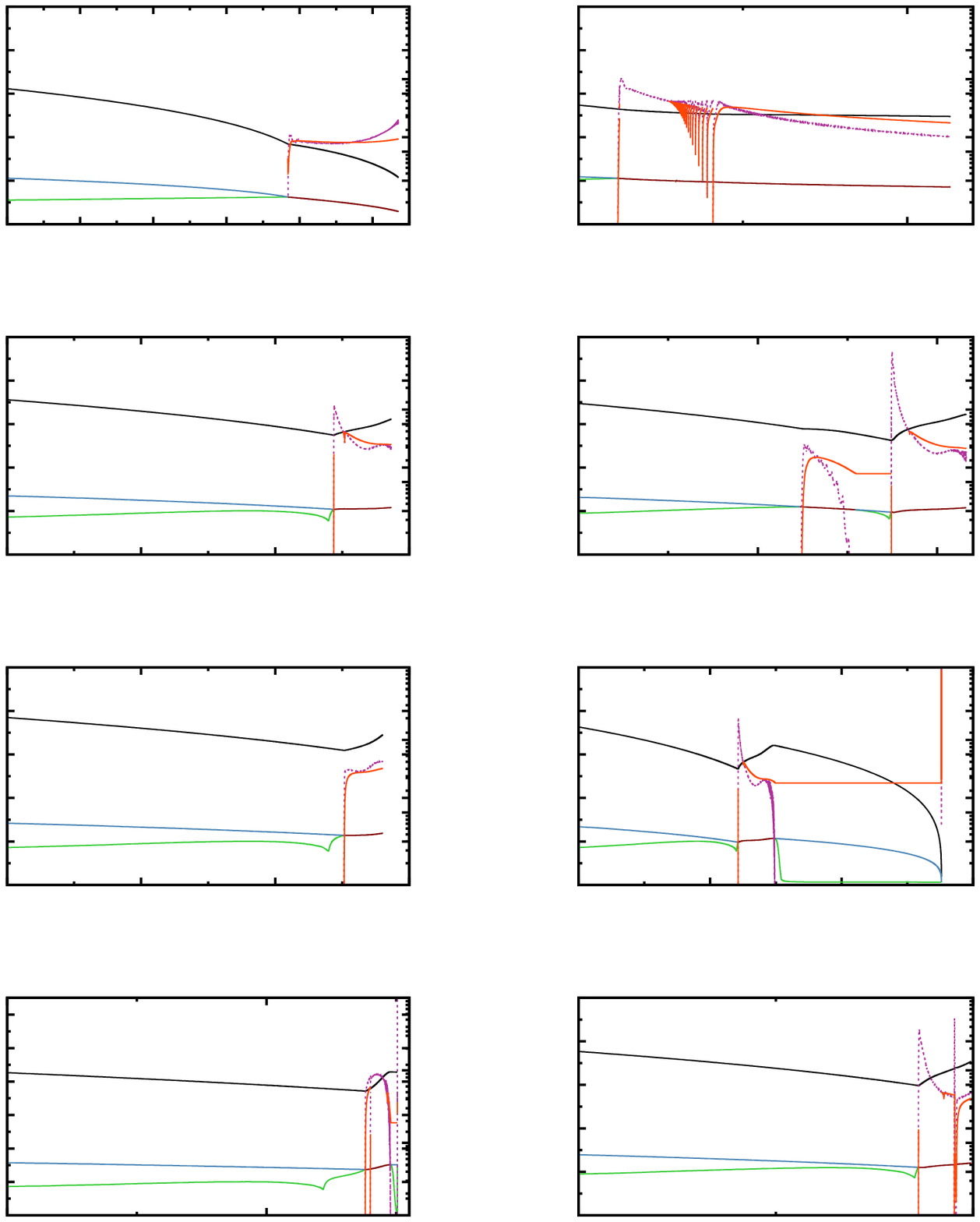}
   \end{subfigure}
   \begin{center}
   \begin{subfigure}[b]{1.0\textwidth}
   \input{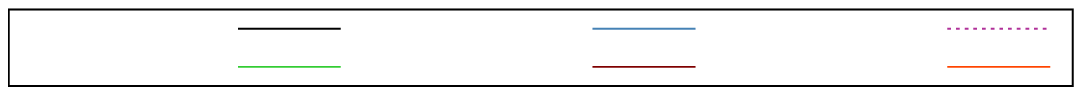}
   \end{subfigure}   
   \end{center}
   \normalsize
      \caption{Orbital parameters and mass transfer rates over time after HeZAMS of various systems studied in this paper. Radii are mapped to the left hand y-axis of each diagram, mass transfer rates to the right. The seven-digit numbers in each diagram denote the identity of the represented system as defined by its initial parameters. The first two digits are equal to the mass of the donor star divided by $0.1\,\text{M}_\odot$, the second two digits are equal to the mass of the WD divided by $0.1\,\text{M}_\odot$, and the last four digits are equal to the initial orbital period divided by $0.001\,\text{d}$. $[\dot{M}]$ is the time-averaged mass transfer rate, $R_1$ the radius of the donor star, $R_\text{Roche,1}$ the radius of the donor star roche lobe. RLOF 1 denotes times when $R_1 \geq R_\text{Roche,1}$ and, consequently, the donor star experiences Roche lobe overflow.}
         \label{fig:multiroche}
\end{figure*}

\begin{figure*}
   \centering
   \small
   \begin{subfigure}[t]{1.0\textwidth}
   \input{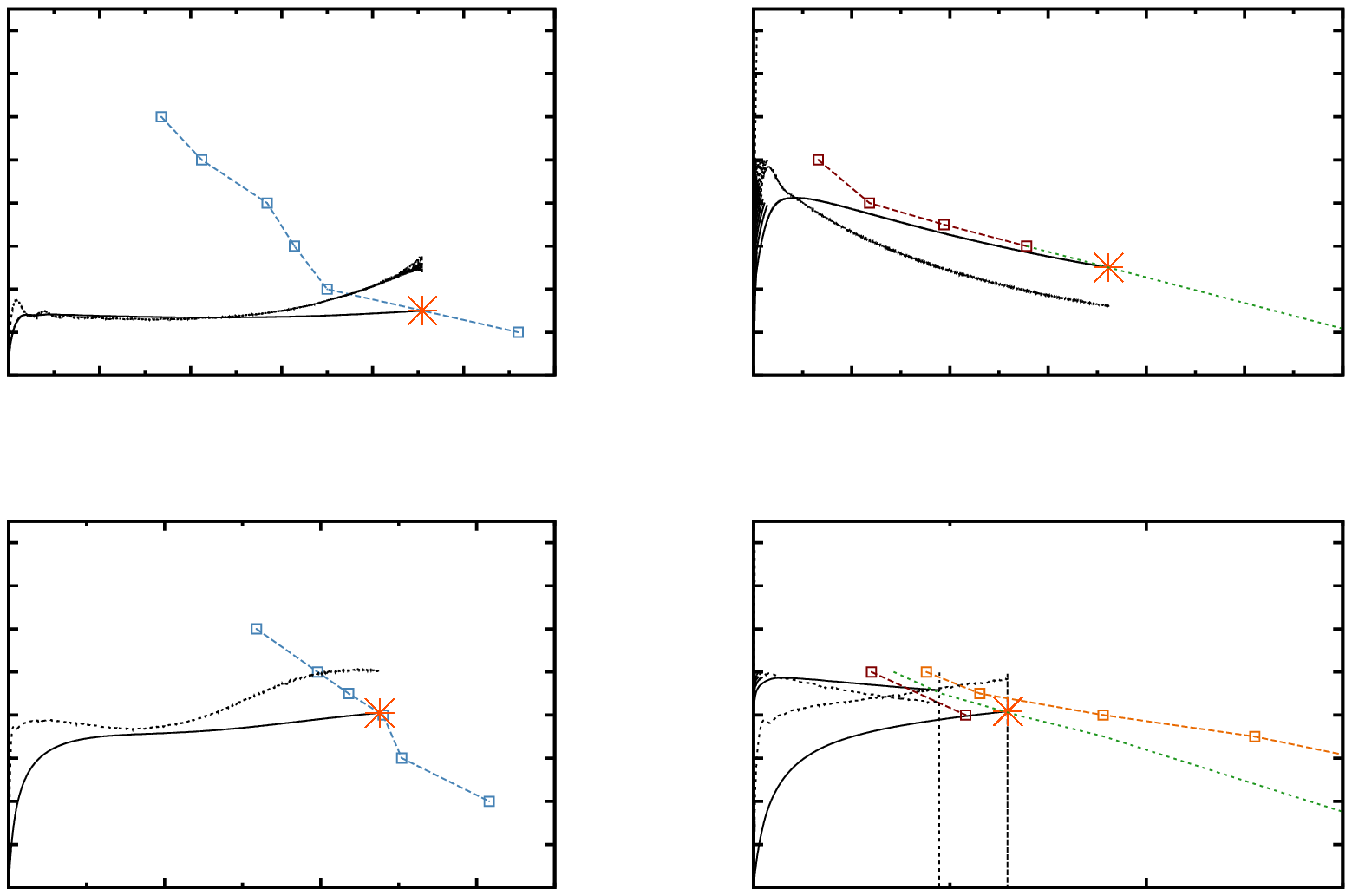}
   \end{subfigure}
   \vspace{-3\baselineskip}
   \begin{center}
   \begin{subfigure}[b]{1.0\textwidth}
   \input{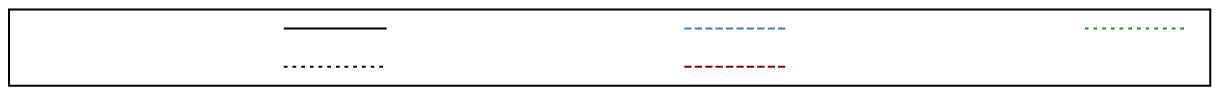}
   \end{subfigure}   
   \end{center}
   \normalsize
   \vspace{-2\baselineskip}
      \caption{Tracks of selected systems in the $\dot{M}-M_\text{Shell}$ phase space. $\dot{M}$ is, as before, the mass transfer rate and $[\dot{M}]$ the time-averaged mass transfer rate. $\dot{M}_c (M_\text{Shell})$ is the detonation line for initially cold models and $\dot{M}_h (M_\text{Shell})$  for initially hot models (see Sec.~\ref{sec:methods}). The orange star represents the state of the system at the moment of detonation. The detonation lines shown in plots a) and c) are appropriate for each WD CO core mass at the time of detonation. In systems b) and d), the given detonation lines did not cover the necessary parameter space and had to be extrapolated to lower mass transfer rates and interpolated with respect to the WD's CO core mass. The resulting inter- and extrapolations are given by $\dot{M}_{h,i/e} (M_\text{Shell})$. In systems b) and d), mass transfer crosses into the nova regime, which allows for a change in mass of the CO core. The final WD's CO core mass of system b) is $0.8~\text{M}_\odot$, that of system d) is $1.059~\text{M}_\odot$. As in Fig.~\ref{fig:multiroche}, the labels identify each system by means of their initial parameters.}
         \label{fig:multidet}
\end{figure*}

\emph{Steady accretors:} These systems experience a single episode of mass transfer until detonation occurs or until the mass transfer stops as the donor star becomes a WD. One implication of this mode of mass transfer is that the donor star will still be on the helium main sequence at detonation, if detonation occurs. These systems tend to have relatively low mass transfer rates ($\dot{M} \approx 1 \cdot 10^{-8} \, \text{M}_\odot/\text{yr}$), and hence relatively high He masses at detonation ($M_\text{He}>0.15 \, \text{M}_\odot$). Steady accumulation ($\eta=1$) will lead to fairly predictable outcomes, as is discussed in Sec.~\ref{sec:WD0.8}.
An example of this system type is 05080050. As shown in Fig.~\ref{fig:multiroche}a, mass transfer starts while the donor star is still on the helium main sequence. It proceeds at a relatively low rate, growing only as a result of angular momentum transport to the accretor until the detonation conditions are met, as shown in Fig.~\ref{fig:multidet}a, and then the WD detonates.

\emph{Non-steady accretors:} These systems cross the boundary between $\eta = 1$ and $\eta = 0$ (see Fig.~\ref{fig:acceff_prescription}) multiple times during their evolution. Since $\eta=0$ means that the transferred He is ejected by helium deflagrations, significant amounts of matter can be removed from the system through multiple  crossings. The donor mass and the mass transfer rate decrease in the subsequent evolution. This can lead, in some high-mass-donor systems to otherwise unexpectedly massive helium shells at detonation. Usually, as explained in Sec.~\ref{sec:donor}, massive donor stars would imply high mass transfer rates, which then lead to low helium shell masses at detonation (see Sec.~\ref{sec:methods} and Fig.~\ref{fig:detlines}). Non-steady accretion can occur for any donor mass, but in low mass donors, it will invariably remove such a  large  amount of helium from the donor that the helium remaining on the donor afterwards will be insufficient to cause a detonation on the WD. Non-steady accretion occurs exclusively in  case  BA-systems\footnote{In mass transfer from hydrogen-rich stars, one differentiates between case A and case B mass transfer. In case A systems, mass transfer occurs before the end of central hydrogen burning and afterwards in case B systems. In the same vein, case BA systems begin their first mass transfer phase before the end of central helium burning and case BB systems afterwards.}. Non-steady accretors are therefore a special class among the systems of this study. One such system is 10080035. As shown in Fig.~\ref{fig:multiroche}b, the critical mass transfer rate for deflagrations is crossed multiple times while the donor star is still proceeding through helium main sequence evolution, each mass ejection widening the orbit and reducing the mass transfer rate until the rate is low enough for steady accumulation of helium on the accretor. Fig.~\ref{fig:multidet}b shows that the detonation line is then crossed, and the WD detonates.

\emph{Single deflagration:} These are systems that cross the boundary from $\eta = 0$ to $\eta = 1$ exactly once.
Systems of this kind invariably cross into the regime of $ 0 < \eta < 1$ initially. As in non-steady accretors, mass transfer from the donor is very efficient, but instead the mass is partially or completely added to the mass of the accretor's CO core.
Systems of this type usually exhibit very small helium shell masses at detonation, which are due to the accretor being heated by a deflagration early in its evolution and which makes them prime precursor candidates for ordinary-looking SNe Ia.
One such system is 08080055, shown in Fig.~\ref{fig:multiroche}c. Here, initially high mass transfer rates are continually reduced by the increase of the donor's nuclear timescale following mass loss and angular momentum transfer to the accretor. Eventually the mass transfer rate decreases sufficiently for steady accumulation of helium to become possible, and the WD detonates.

Single deflagration systems and non-steady accretors enter a phase of continual mass loss by crossing the $ [\dot{M}] > \dot{M}_{\text{WK,max}}$-boundary as described in Sec.~\ref{sec:methods}.

\emph{Biphasic:}
Biphasic systems, which occur exclusively  in systems with mass ratios close to unity, are characterized by a limited phase of RLOF during the helium main sequence lifetime of the donor where the conditions for detonation are not met. The donor then contracts, ending mass transfer before initiating shell helium burning and expanding rapidly, which results in extremely high mass transfer rates, eventually reducing the mass transfer rates sufficiently for detonation to occur. An example of this kind of system is shown in Fig.~\ref{fig:multiroche}d.

\subsubsection{Systems with $M_\text{WD}=0.8~\text{M}_\odot$} \label{sec:WD0.8}

\begin{figure*}
   \centering
   \small
   \begin{subfigure}[t]{1.0\textwidth}
   \input{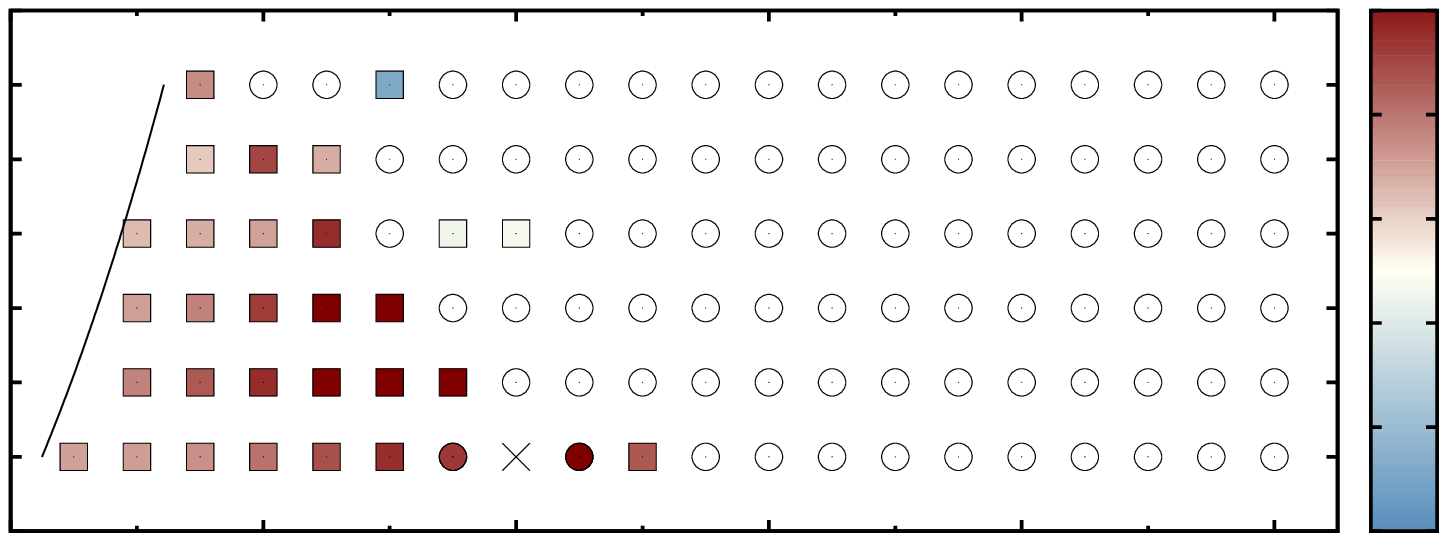}
   \end{subfigure}
   \begin{subfigure}[]{1.0\textwidth}
   \input{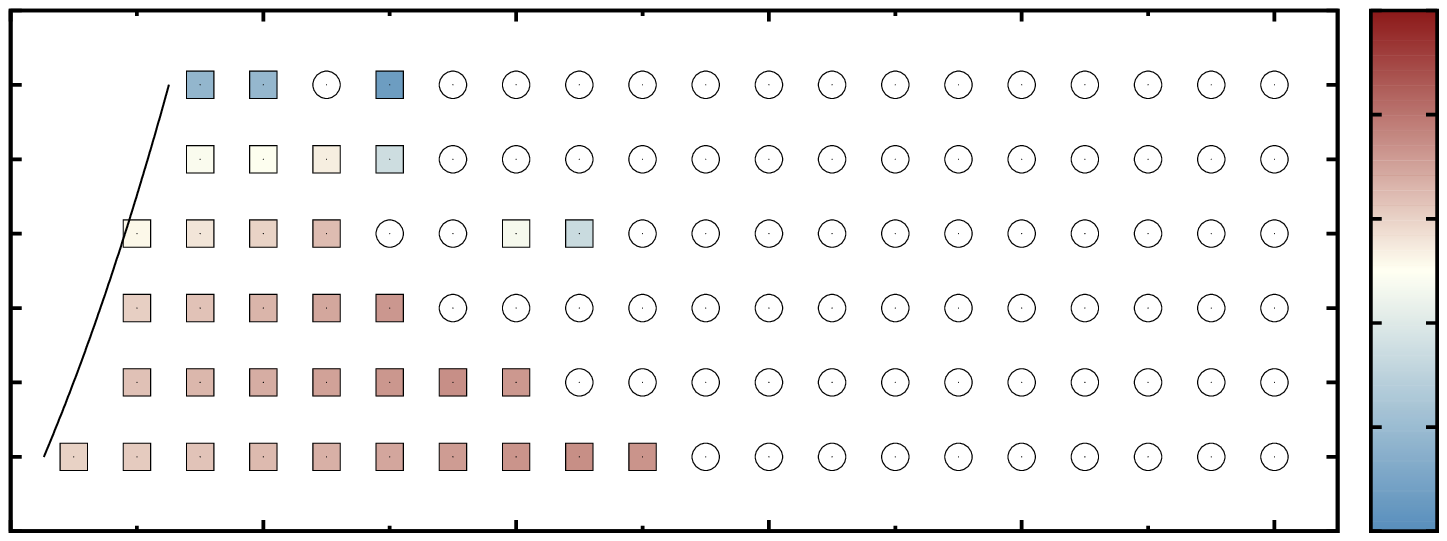}
   \end{subfigure}
   \begin{subfigure}[b]{1.0\textwidth}
   \input{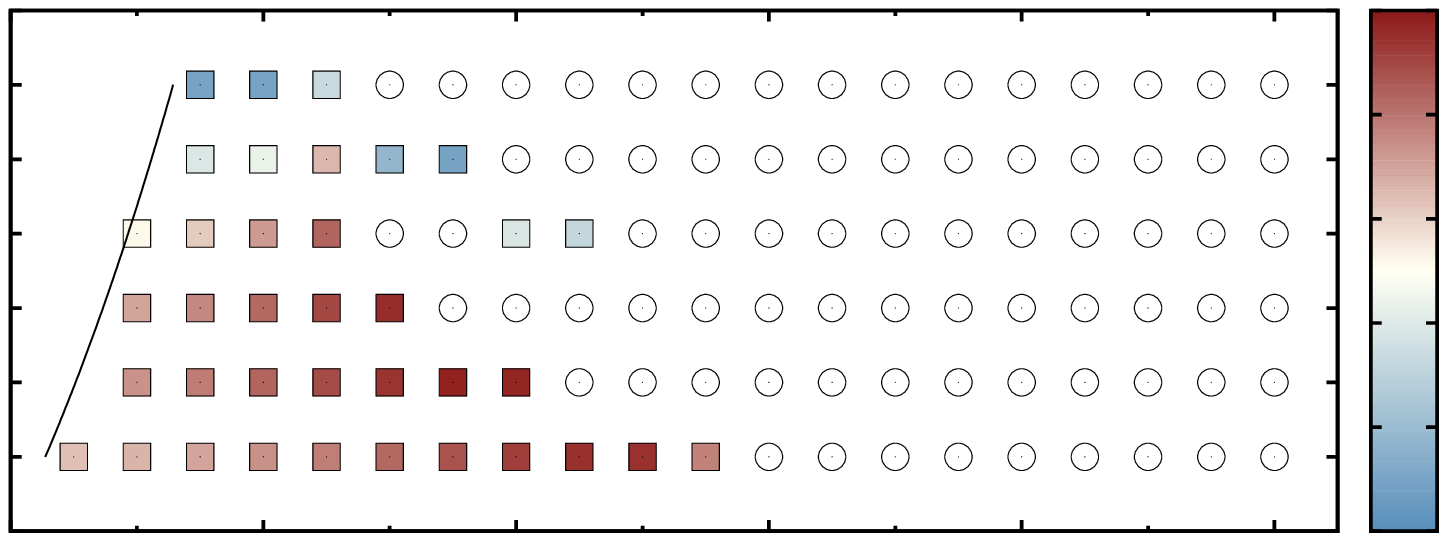}
   \end{subfigure}
   \normalsize
      \caption{Positions of individual systems in the initial parameter space. $M_\text{d,i}$ is the initial mass of the mass losing helium star. Filled squares indicate systems that experience detonations at some point in their evolution, circles systems that do not. Crosses denote numerically unstable systems whose evolution could not be tracked to the point of  detonation or to a time when the binary is unable to produce a detonation due to He-depletion in the donor. The color scale (applicable to detonating systems only) indicates the amount of helium present ($M_\text{He}$) on the surface of the white dwarf at the time of detonation. Systems located on the black line  fill their Roche lobes from the beginning of the simulation. The identity of each system shown in Fig.~\ref{fig:multiroche} is indicated.}
         \label{fig:grids}
\end{figure*}

Fig.~\ref{fig:grids}a shows the final helium shell masses of CO WDs in systems with $M_\text{WD}=0.8\,\text{M}_\odot$.
Below $M_\text{Donor} = 0.7\,\text{M}_\odot$, steady accretion dominates (see Fig.~\ref{fig:multiroche}a for example). The timescale for orbital angular momentum loss decreases as the orbit gets smaller, and the mass transfer rate increases accordingly. This leads to a lower helium shell mass for a shorter orbital period for a given donor mass. As is seen in Fig.~\ref{fig:multiroche}a, stars with masses of less than $0.7\,\text{M}_\odot$ expand slightly during their helium main sequence. Longer initial periods therefore mean that the donor fills its Roche lobe when both stellar and Roche-lobe radii are larger, thus only allowing relatively weaker enhancement of the mass transfer rate due to emission of GWR, in turn leading to larger helium shell masses at detonation.

The decrease in the helium shell mass at the upper period boundary of the detonation parameter space is expected. Longer initial periods mean that RLOF will happen closer to the end of the donor star's main sequence lifetime.  Towards the end of the donor's main sequence evolution, mass transfer is mainly driven by the shorter nuclear timescale of the star, inducing higher mass transfer rates. This results in smaller He-shell masses at detonation (see Fig.~\ref{fig:detlines}). The mechanism of this decrease is applicable to all systems in this class.

It is also evident that the mass of the helium shell necessary to detonate a CO WD of a given mass decreases with increasing donor star mass. This is consistent with expectations that the mass transfer rate becomes higher as the nuclear timescale becomes shorter.

System 10080035 ($M_\text{Donor} = 1.0\,\text{M}_\odot$ $M_\text{WD} = 0.8\,\text{M}_\odot$, $P = 0.035\,\text{d}$) exhibits non-steady accretion (Figs.~\ref{fig:multiroche}b and \ref{fig:multidet}b). This system first experiences a high mass transfer rate, driven by its relatively short timescales of nuclear burning and GWR. The mass ejected from the system is mostly lost via small-scale ejections. With the orbit widening under the influence of mass exchange, the radius of the Roche lobe exceeds the stellar radius and the mass transfer stops. The donor continues to contract for a while thereafter, but expands again while the Roche lobe continues to contract owing to GWR. Once the star fills its Roche lobe again, depending on the star's current mass, the mass transfer rate may again exceed $\dot{M}_{\text{WK,max}}$, inducing a deflagration and ejecting the helium accumulated on the WD up to that point. This in turn inflates the donor star's Roche lobe and prompts mass transfer to cease, until the donor fills its Roche lobe once more.
Repeated mass ejections lead to the loss of $0.16\,\text{M}_\odot$ in this system.

The two systems 08080055 and 08080060 are separated from the main body of the detonation zone (08080060 depicted in Fig.~\ref{fig:multiroche}e). This seems to be caused by an interplay between the radius increase of the donor star and the loss of angular momentum via GWR. Systems with longer and shorter periods (namely 08080055 and 08080070, Figs.~\ref{fig:multiroche}f and \ref{fig:multiroche}g, respectively) either exhibit too high a mass transfer rate or fail to accumulate sufficient mass in order to initiate a detonation. This shows that He detonation may happen even in parts of the parameter space which is unconnected to the main detonation zone.

This particular grid contains two systems, denoted by filled circles, which became numerically unstable after approaching their detonation line to within 1\% of the mass needed for detonation. Most likely these systems would eventually detonate, but we are unable to exactly calculate, within our methodology, the necessary amount of mass. In these cases, the color shading indicates the amount of mass that was accumulated at the point where the model became unstable, with the understanding that the amount of helium present on the WD, were it to detonate, would be slightly larger than  indicated.
\subsubsection{Systems with $M_\text{WD}=0.9~\text{M}_\odot$}
Fig.~\ref{fig:grids}b shows the final helium shell masses of CO WDs in systems with $M_\text{WD}=0.9\,\text{M}_\odot$. As with systems containing WDs of masses of $0.8\,\text{M}_\odot$, almost all detonating systems containing donor stars of $M_\text{Donor} \leq 0.9\,\text{M}_\odot$ undergo steady accretion; system 10090040 is of the non-steady type and Systems 09090050 (discussed below) and 10090050 are biphasic.

System 09090050 (see Fig.~\ref{fig:multiroche}d) is an interesting case of biphasic mass transfer. Case BA mass transfer  occurs initially, but the star contracts in response to high mass loss rates, then expands as a result of core helium depletion, initiating case BB mass transfer. The initial mass transfer rate in this second mass transfer episode is high, but decreases owing to the increase in the orbital separation, induced by mass transfer to the, then, more massive accretor, which then detonates.

System 10090050 is separated from the rest of the detonating systems. Both of these systems are biphasic, but in 10090045 -- owing to the  smaller orbital separation at the start of the BB phase -- a deflagration is triggered. This leads to a premature ejection of the helium envelope of the white dwarf, inhibiting detonation.

\subsubsection{Systems with $M_\text{WD}=1.0~\text{M}_\odot$}

Fig.~\ref{fig:grids}c depicts the final helium shell masses in the systems containing a 1.0\,$\text{M}_\odot$ WD. All of these systems except 0910005, 0910006, 0810006, and 0810007 feature steady case BA accretion. System 0910005 undergoes two distinct phases of mass transfer and the rest of the mentioned systems are of the single deflagration type. The unexpected \citep[compare to][]{WJH2013} extension of the detonation zone towards longer initial periods is due to the WD requiring less helium to be accumulated for detonation once it has been heated through accretion at high mass transfer rates. Other features of this data set are similar to those containing lower mass WDs.

We note that the area of all these parameter spaces where detonation follows a single episode of case BA mass transfer corresponds well to the contour found by \citet{WJH2013}.

\subsubsection{Pre-detonation conditions}

\begin{figure}
   \centering
   \small
   \input{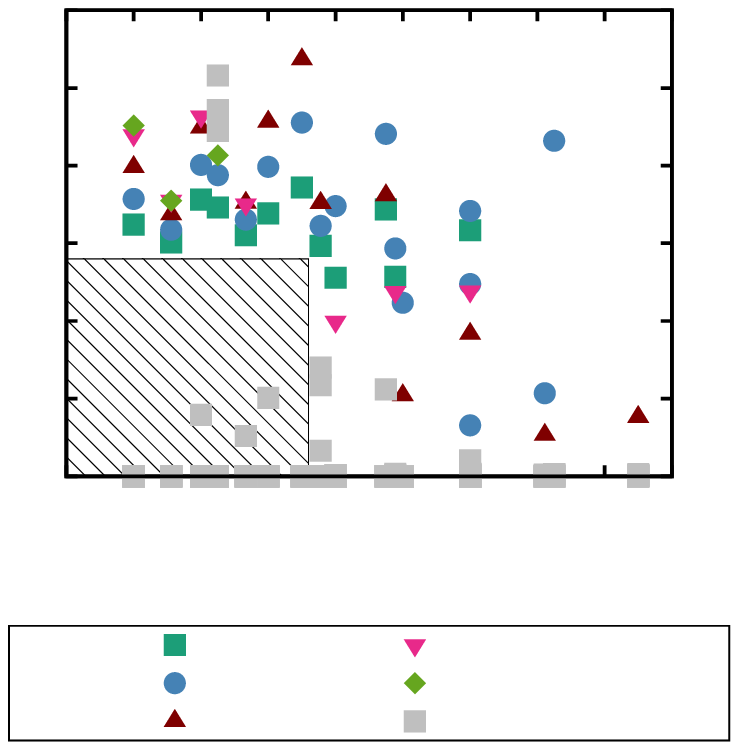}
   \normalsize
      \caption{Final He-shell masses of all simulated systems over $M_1/M_2$. Gray squares indicate systems that do not experience detonation during their evolution. Colored points indicate systems that do produce a detonation and their initial period. It is evident that detonation seems to be excluded for small values of $M_1/M_2$ (hatched area). The exclusion is due to the decreasing gradient of the detonation line with decreasing $\dot{M}$, while lower mass WDs need to accumulate a larger mass of helium in general.}
         \label{fig:mr-he}
\end{figure}

Fig.~\ref{fig:mr-he} shows the final mass of the accumulated helium shell of our models over the initial mass ratio (here defined as $q = M_1/M_2$). Helium shell masses of less than about $0.14\,\text{M}_\odot$ are realized only at mass ratios higher than $q=0.8$, i.e., there is a marked absence of systems with both small final helium shell mass and small mass ratios, roughly indicated by the hatched area,  because  lighter WDs require a larger buildup of helium \citep[see][]{SB2009,IT1991,WK2011} and  the higher accretion rates from a higher mass donor star will cause stable burning on the WD, lowering the helium threshold for detonation. 
The apparent bias of longer initial periods towards larger final He-shell masses is mainly due to the fact that only lower mass donors will initiate a detonation at longer periods.

\begin{figure}
   \centering
   \small
   \input{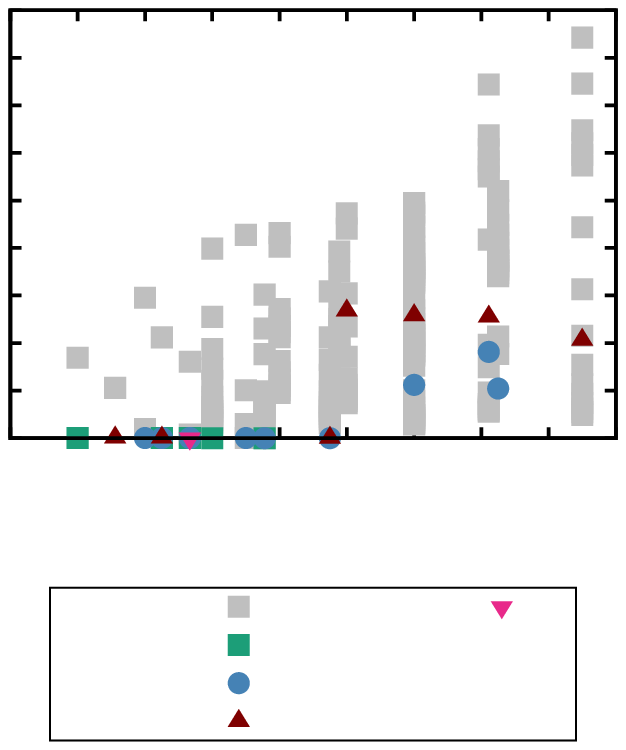}
   \normalsize
      \caption{Cumulative mass loss from systems over $q$, with systems that experience no mass loss not included. Gray squares indicate systems that experience mass loss but no detonation. Colored systems, with color indicating initial orbital period, do experience detonation. It is immediately evident that systems with low $q$ only experience significant mass loss if detonation does not occur.}
         \label{fig:mr-loss}
\end{figure}

Fig.~\ref{fig:mr-loss} shows the total mass lost by a system as a function of $q$. No system with $q\leq0.78$ undergoes any appreciable mass loss before detonation because mass donors will, after a short phase of expansion, contract to become a white dwarf.

Systems with $q\geq0.78$, however, do lose appreciable amounts of helium before detonation. This is due to higher mass donors transferring mass to the accretor, if RLOF happens during their helium main sequence, at higher mass transfer rates corresponding to their relatively shorter nuclear timescale. Because the transfer rates are high enough to trigger a nova-cycle, the transferred mass is then lost from the system. With their initial mass reduced, and, consequently, their nuclear timescale increased, expected mass transfer rates will be lower. This means that these initially more massive donors can still contribute sufficient amounts of helium at sufficiently low mass transfer rates to lead to detonation.

At this point it should be mentioned that \cite{MW2015} have found evidence for a correlation between SNe Ia and observable circumstellar matter. However, since the mass ejections in our systems generally happen several Myr before detonation, the ejected matter will generally have dispersed sufficiently at the point of detonation to no longer be detectable, though we do not discount the possibility of circumstellar medium being generated by processes which cannot be resolved by our methodology.

\subsubsection{Final helium mass distribution and implications for ordinary SNe Ia}

\begin{table}
\centering
\caption{Parameters of hot WDs with accreted helium exhibiting normal type Ia spectra after detonation according to \cite{WK2011}. 
$M_{WD}$ is, as before, the mass of the WD's CO core at the time of detonation and $M_{He}$ is the mass of the accumulated helium layer.}
\label{tab:Iadata}
\begin{tabular}{l l}
\hline\hline
$M_{WD}$ & $M_{He}$ \\
\hline
0.8 & 0.097 \\
0.9 & 0.055 \\
1.0 & 0.0445 \\
1.1 & 0.027 \\
\hline \\
\end{tabular}
\end{table}

\begin{figure} 
        \centering
        \small
       \input{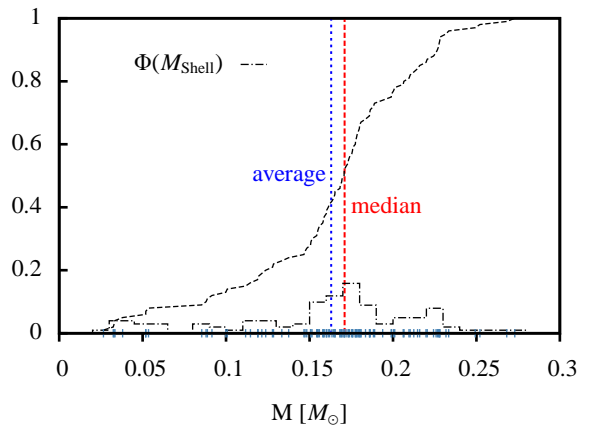}
        \normalsize
        \caption{Cumulative probability distribution of accumulated He-shell masses prior to detonation. The probability of a system having a given set of initial parameters at the HeZAMS is assumed to be equal for the entire initial parameter space. The blue tics at the bottom of the graph show the helium shell mass of all individual data sets. The vertical lines indicate the median and average He-shell masses at detonation.}
        \label{fig:cumulative}
\end{figure}

The cumulative probability for detonations involving less than a specified amount of helium based on the grids produced in this study is depicted in Fig.~\ref{fig:cumulative}. It is immediately evident that in our parameter space there is a clear bias toward helium shell masses larger than $0.15\,\text{M}_\odot$. This, again, is due to the fact that detonations induced through steady accretion in systems with small mass ratios are clearly favored. The median He-shell mass at detonation is found at $0.171\,\text{M}_\odot$ and the average mass at $0.163\,\text{M}_\odot$.

\begin{figure} 
        \centering
        \small
       \input{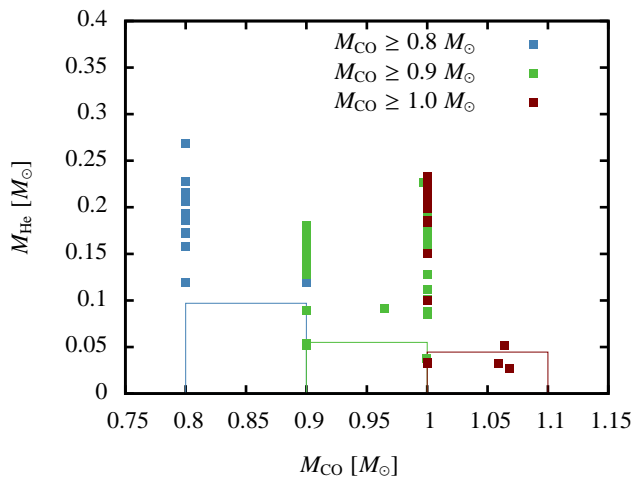}
        \normalsize
        \caption{Final He-Shell masses of detonating WDs with respect to their core mass. The colored outlines represent areas in which systems are considered to fulfill the conditions set in Table~\ref{tab:Iadata} for producing normal SN Ia-like spectra  at detonation. Each outline is valid only for dots of the same color.}
        \label{fig:CO-Shell}
\end{figure}

\cite{WK2011} computed spectra for different WD masses and a range of final helium shell masses.
The positions of our models in the stated parameter space is visualized in Fig.~\ref{fig:CO-Shell}.
In Tab.~\ref{tab:Iadata}, we list the systems that satisfy the low He mass condition to produce ordinary SNe Ia. 
There is a significant bias towards mass ratios of $q \geq 1$ and initial orbital periods smaller than $0.06$\,d for the formation
of these systems. 
All of the steady case BA systems accumulate too much helium such that producing an ordinary Type Ia spectrum from an initially cold WD is not likely. Only accretion onto hot WDs, which can be produced by steady case BB mass transfer or novae, would lead to final helium shell masses consistent with  normal SNe Ia spectra.
The fraction of these low helium shell systems of all detonating systems is only about $0.1$. 
We conclude that short-period binary systems consisting of a He star and a CO WD in the considered mass range cannot be considered a major channel towards ordinary SNe Ia, but they might contribute significant to the rate of peculiar SNe Ia. 

\begin{table}
\centering
\caption{Systems exhibiting low enough mass accumulation at detonation for which ordinary SN Ia spectra are expected.}
\label{tab:Iares}
\begin{tabular}{l l l l}
\hline\hline
SID & $M_{\text{WD,f}}$ & $M_{\text{He,f}}$ & Acc. Type\\
\hline 
10090050 & 1.09484 & 0.0265 & biphasic \\
09100055 & 1.09176 & 0.0324 & biphasic\\
10100040 & 1.03293 & 0.0329 & non-steady\\
10100035 & 1.0334 & 0.0334 & non-steady\\
10080050 & 1.03758 & 0.0380 & steady\\
10090035 & 0.9518 & 0.0518 & non-steady\\
10090040 & 0.95350 & 0.0535 & non-steady \\
\hline
\end{tabular}
\end{table}

\subsubsection{Properties of donors at detonation}

\begin{figure} 
        \centering
        \small
       \input{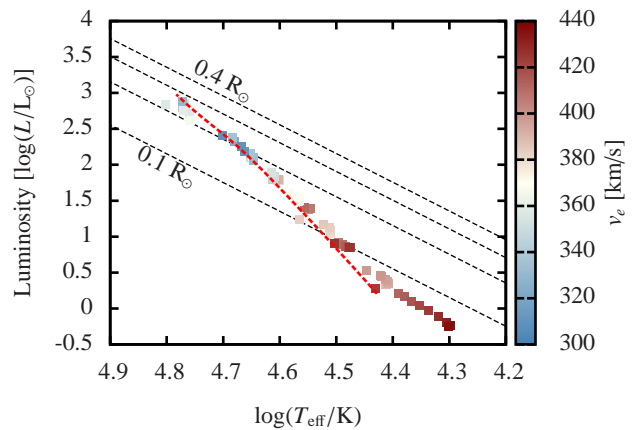}
        \normalsize
        \caption{Positions of the donor stars in the HR diagram at the point of detonation of the white dwarf. Color shading indicates the final orbital velocity of the donor. The dashed red line corresponds to the zero age main sequence for helium stars of masses between $0.325~\text{M}_\odot$ and $1.5~\text{M}_\odot$.}
        \label{fig:endhrs}
\end{figure}

The donor star would be unbound from the binary system once the detonation disrupts the WD.  Fig.~\ref{fig:endhrs} shows the positions of the donor stars in the HR diagram. The stars lie in close proximity to the helium main sequence. Fig.~\ref{fig:endhrs} also reveals a clear bias for brighter donors to have smaller final orbital velocities, which have been found to lie between $\sim 300~\text{km/s}$ and $\sim 440~\text{km/s}$.
A similar study was conducted by \cite{WH2009}, though using a different methodology. Owing to the differences in methodology and range of the initial parameters, the results of this study cannot easily be quantitatively compared with these previous results, but a qualitative comparison of the expected final orbital velocities can be used as a sanity check on both studies. While we can therefore neither confirm nor disprove the distributions calculated in \cite{WH2009}, we can at least confirm that in the relevant portion of the parameter space, the final orbital velocities of the donor star correspond closely to the values found in the present study.
It should also be noted that the observed runaway helium star found by \cite{GFZ2015} exhibits a galactic rest frame velocity higher than found in our sample. Since that particular runaway star has already left the galactic plane and has therefore lost a significant fraction of its initial velocity, its current state is not explained by these calculations. \\
Recently, progress on the question of the immediate pre-detonation state of SN Iax progenitors has been achieved with the observation of one such object by \cite{MJS2014}, who have confirmed that the progenitor of SN 2012Z was a bright, blue source. The exact nature of this source is still being debated. Data seems to point to either a main sequence (hydrogen) star of $M\sim18.5~\text{M}_\odot$, an early stage blue supergiant of $M\sim11.0~\text{M}_\odot$, or a late stage blue supergiant of $M\sim7.5~\text{M}_\odot$. However, a helium giant of $M\sim1.2~\text{M}_\odot$ cannot be  excluded. Measuring the maximum luminosity at shock breakout, \cite{YMK2015} determined the radius of the source to have been $\sim2.0~\text{R}_\odot$. Assuming that the source is indeed a helium star, this radius measurement would indicate an extended envelope. While, at the time of writing, observation supports the notion that the above mentioned source was part of an interacting binary, robust observational confirmation of these data will require the SN itself to have faded. Discounting the possibility of the radius measurement being influenced by the existence of an accretion disc, it is likely that this particular progenitor is unlike the short-period WD+He star binaries studied here and that at least either the detonation mechanism or, if the progenitor is indeed a binary component, the mass accretion or mass transfer mechanism must be of a different nature. However, a recent effort \citep{FDJ2015} has failed to identify the progenitor of the type Iax supernova SN 2014dt, but has   excluded sources that are very similar to the progenitor of SN 2012Z. This indicates that  a variety of progenitor systems may be capable of producing SN Iax events, which, at present, do not seem to exclude WD+He star systems in general and double detonation scenarios in particular.

\section{Conclusions} \label{sec:conclusions}
We conducted an extensive survey of the parameter space of very short-period He-star CO WD systems, varying initial donor mass, WD mass, and initial orbital period. Parameterizing detonation conditions and mass accretion efficiency according to \cite{WK2011} and \cite{HK2004}, we found that accumulated helium shell masses in the single degenerate double-detonation scenario  are strongly biased toward values larger than $\approx 0.1\,\text{M}_\odot$. Using the final helium shell mass as a yardstick as to the likelihood of resulting spectra resembling normal type SNe Ia, we infer that these spectra might be produced by about 10\% of the systems in our sample that result in a detonation. With the rest of our sample we only expect peculiar SNe Ia-like events, which, according to \cite{WK2011}, might include SNe Iax.\\
We conclude that the systems considered in this study are unlikely to account for more than a small fraction of the normal SN Ia rate. The obvious reason for this conclusion is that, if this were  the main channel of SN Ia production, a much larger ratio than the reported value of 5-30\% of SN Iax to normal SN Ia would be expected to be observed \citep{MJS2014}.\\
We predict that the average WD in our sample will accumulate $0.163\,\text{M}_\odot$ of non-processed helium before detonating. The apparent absence of systems of this configuration in nature could point towards non-negligible contributions from physical processes not included in this study, such as the effects of rotation and/or magnetic fields, as well as mass accretion dynamics (i.e., the existence of an accretion disk and non-uniform distribution of the accreted material). Quantitative issues might arise from the sparsity of previous model predictions concerning ignition conditions for a CO WD with respect to its mass, mass transfer rate, and temperature, as well as the predicted mass retention efficiency.
The possibility of helium accretors of the considered mass range being a significant contributor to supernova events exhibiting peculiar spectra merits further investigation, though the marked absence of observed events featuring amounts of helium as large as the ones predicted by this study casts doubt on the assumption of systems of the type considered here being a significant contributor to SNe Iax rates.\\
Furthermore, the question of how much variation in the spectrum could be expected for slight variations in the WD-mass and temperature would be a promising line of inquiry. Also, significant changes in the predicted outcomes of these systems would be expected depending on the state of the WD at the beginning of interaction with the He-rich companion due to the variation in the accreted helium mass required for detonation depending on the initial temperature of the accretor.  In all likelihood, a more detailed understanding of the evolution of a degenerate component in a common-envelope system would be advantageous.

\begin{acknowledgements}
P. N. would like to extend his gratitude to Dr. Zhengwei Liu and Dr. Thomas Tauris for useful discussions. This research was supported by the Deutsche Forschungsgemeinschaft (DFG). Grant No. Yo 194/1-1.
\end{acknowledgements}

\bibliographystyle{aa}
\bibliography{AA201527845.bib}{}
\end{document}